\documentclass[apj]{emulateapj}
\setkeys{Gin}{width=0.45\textwidth}

\newcommand{\kpc}{{\rm kpc}}

\newcommand{\hmpc}{\ifmmode{h^{-1}\,\hbox{Mpc}}\else{$h^{-1}$\thinspace Mpc}\fi}

\newcommand{\kms}{\ifmmode{\,\hbox{km\,s}^{-1}}\else {\rm\,km\,s$^{-1}$}\fi}
\newcommand{\msun}{{\rm\,M_\odot}}

\begin{document}
\title{Dark Matter Sub-Halo Counts via Star Stream Crossings} %\altaffilmark{1}}
\shorttitle{Star Stream Gaps}
\shortauthors{Carlberg}
\author{R. G. Carlberg\altaffilmark{1} }
\altaffiltext{1}{Department of Astronomy and Astrophysics, University of Toronto, Toronto, ON M5S 3H4, Canada carlberg@astro.utoronto.ca }
\begin{abstract}
Dark matter sub-halos create gaps in the stellar streams orbiting in the halos of galaxies. We evaluate the sub-halo stream crossing integral with the guidance of simulations to find that the linear rate of gap creation, $\mathcal{R}_\cup$, in a typical Cold Dark Matter (CDM) galactic halo at 100 kpc is $\mathcal{R}_\cup \simeq  0.0066{\hat{M}_8}^{-0.35} \, {\rm kpc}^{-1} {\rm Gyr}^{-1}$, where $\hat{M}_{8} (\equiv \hat{M}/10^{8} \msun$) is the minimum mass halo that creates a visible gap. The relation can be recast entirely in terms of observables, as $\mathcal{R}_\cup \simeq 0.059 w^{-0.85}\, {\rm kpc}^{-1} {\rm Gyr}^{-1}$, for $w$ in kpc, normalized at 100 kpc.  Using published data, the density of gaps is estimated for M31's NW stream and the Milky Way Pal~5 stream, Orphan stream, and Eastern Banded Structure. The estimates of the rates of gap creation have errors of 50\% or more due to uncertain dynamical ages and the relatively noisy stream density measurements. The gap rate-width data are in good agreement with the CDM predicted relation.  The high density of gaps in the narrow streams require a total halo population of about $10^5$ sub-halos above a minimum mass of $10^5\msun$. 
\end{abstract}

\section{INTRODUCTION}
\nobreak
The dark matter halos in which galaxies are embedded are satisfactorily modeled as smooth quasi-isothermal spheroids, such as Hernquist (1990) or Navarro-Frenk-White (NFW, 1997) functions, as described in authoritative texts such as \citet{BM:98} and \citet{BT:08}. Such mass models, with addition of the visible stars, gas and dust, can adequately account for most of the observed internal kinematics of galactic systems. On the other hand, N-body simulations of the formation of dark matter halos robustly predict that approaching 10\% of the mass of any galactic dark halo should be in the form of sub-halos with numbers that rise steeply towards lower masses \citep{Klypin:99, Moore:99, Aquarius, VL1}. The baryonic disk and bulge components of a galaxy will lead to sub-halo depletion in the central regions \citep{DOnghia:10} relative to a pure dark matter model. Although galaxies do contain visible dwarf galaxies embedded in dark matter sub-halos, the numbers are far short of the many thousands expected. Hence, it appears that either the predicted sub-halos are very dark, or, substantially not present. 

An accurate census of the relatively low mass completely dark sub-halos remains a significant challenge. In principle the dark matter itself must have some cross-section for interaction with either itself or baryons, which could potentially give rise to annihilation radiation, but so far there is no clear association of gamma-rays with known sub-halos \citep{BH:10,Veritas:11}. The gravitational effects of  the large numbers of sub-halos quickly average with distance. Sub-halos are potentially detectable through micro-lensing of multiple-image background sources \citep{MS:98,DK:02}. The anomalous flux ratios of multiple image gravitational lenses are explained with the addition of at least one moderately massive sub-halo projected onto the central region of the galaxy \citep{Vegetti:10}, but this is not a detection of a vast population. Over a Hubble time sub-halos lead to heating of some $\sim 30 \kms$ \citep{StarStreams}. Such heating is not easily detectable in most of the visible components of a galaxy.  

The ongoing discovery of thin, low mass, stellar streams in galaxies opens up the possibility that the sub-halos can be detected through their gravitational effects on star streams. In particular, sub-halos that pass near or through a star stream create a gap in the stream \citep{Ibata:02, SGV:08,YJH:11, M31_NW:11, Helmi:11}. Detecting gaps in stream pseudo-images is observationally less expensive than measuring stream kinematics. The challenge of the image analysis is that the stars in any image are overwhelmingly in the foreground or background and the signal above the statistical noise at any point in the stream is often fairly low. Detections of gaps and lumps have been claimed with good statistical confidence for the Pal~5 stream \citep{Odenkirchen:03,GD:Pal5}, the 100 kpc NW stream of M31 \citep{M31_NW:11},  the Eastern Banded Structure of the galaxy \citep{G11:EBS} and discussed for the Orphan stream \citep{Newberg:10}.  

The purpose of this paper is to develop a basic analytic prediction of the rate at which sub-halos create visible gaps in star streams as a statistical measure of the numbers of sub-halos present. The relation can be applied to either simulations or observational data. In principle, the density of dark matter sub-halos (and other bound structures) in a galactic halo is simply measured with the number of times they cross a stellar stream to create visible gaps \citep{StarStreams,YJH:11}. The rate at which gaps are created in a stream is the product of the density of sub-halos, the distance between sub-halo and stream which creates a gap, and the velocity at which sub-halos cross the stream. To use this simple measure of sub-halo density we need to allow for the mass spectrum of sub-halos and determine what range of velocities and impact parameters create visible gaps. In this paper we use a set of restricted n-body simulations to help develop a gap creation rate formula. The last step is to compare the rate predictions to the available data on gaps in the star streams in the Milky Way and M31 as a test of the CDM sub-halo predictions.

\section{The Linear Gap Creation Rate}

The rate at which gaps are created in a long thin stellar stream is the rate at which sub-halos of mass $M$ with local volume density $\mathcal{N}(r,M)$ cross the stream, integrated out to the maximum impact parameter,  $b$, which creates a detectable gap. Full n-body simulations of the sub-halo content of halos find that the distribution of sub-halos, $\mathcal{N}(r,M)$, is separable into a radial dependence, $n(r)$, and a mass function, $N(M)$, such that $\mathcal{N}(r,M)=n(r) N(M)$ \citep{Aquarius}. The $N(M)$ function is expected to have a large variation from one galaxy to another \citep{Chen:11}.
The rate per unit length at which sub-halos cross the stream to create gaps is therefore,
\begin{equation}
{\mathcal R}_\cup(r) = \int_M \int_{v_\perp} \int_0^{b_{max}} n(r) N(M) v_\perp f(v_\perp) \pi\, db \, dv_\perp \, dM.
\label{eq_Ratedef}
\end{equation}
The factor of $\pi$ arises from the consideration that at a distance $b$ from the stream, for any point on a  circumference $2\pi b$, half the stars will be headed inward in a velocity distribution without flows. This analysis will assume that the velocity distribution is Gaussian and isotropic, but more general distributions could easily be incorporated.

The gap rate integral, Equation~\ref{eq_Ratedef}, depends on the visibility of gaps with the distance of closest approach, mass and velocity of a sub-halo. These factors can be evaluated using numerical simulations which determine the average outcome of a single sub-halo passing near or through a stream of particles. Here a star stream is idealized as a ring of particles on a circular orbit.  The mass, velocity, angle of approach and impact parameter are all varied. After the sub-halo has passed we measure the height, length and density dip in the perturbed region, from which we develop a model which can predict $b_{max}$ in terms of sub-halo mass and encounter velocity. With this numerical guidance in hand, we will integrate over sub-halo velocities and the mass distribution, yielding the rate in terms of the minimum sub-halo mass, $\hat{M}$, that creates a visible gap. A related calculation gives the gap lengths in terms of sub-halo masses and can be used to create a relation between gap densities and stream width, both of which are measurable in imaging observations.

\begin{figure}
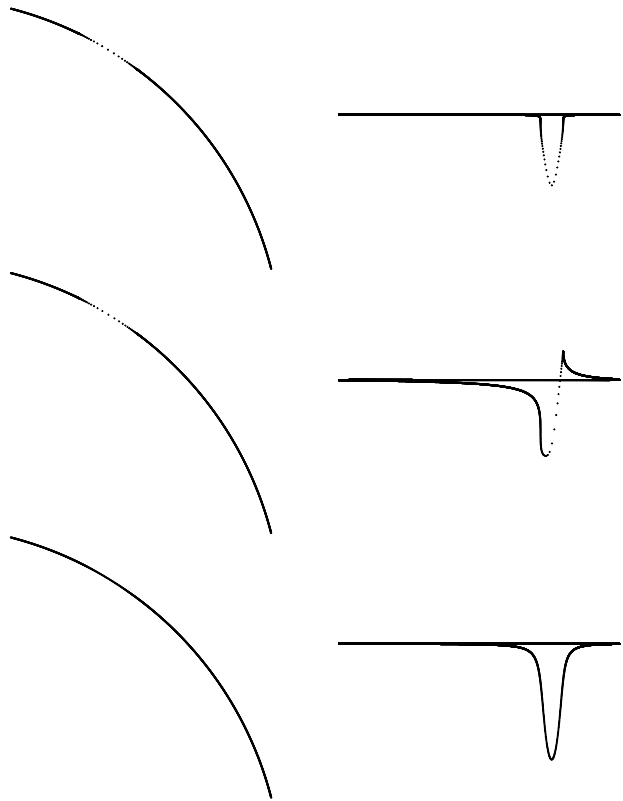

\begin{center}
\includegraphics{Fig1a.eps}
\includegraphics{Fig1b.eps}
\includegraphics{Fig1c.eps}
\end{center}
\caption{The top and side views, left and right, respectively, of rings that a sub-halo of mass $10^8\msun$, with a scale radius 0.83 kpc, passes near. The left column is in boxes of 51 kpc on a side showing the segment of the ring where the sub-halo encounter occurred.  The right column of sub-panels are 132 kpc wide and 0.6 high, that is, vertically stretched a factor of 220.   In all cases the sub-halo motion is radial. The top two rows both have impact parameters of 0.06 kpc, the top being motion parallel to the orbital plane and the middle at an angle of 30\degr\ to the plane. In the bottom row the closest separation is 1.5 kpc above the plane.  The particles are shown at 6.92 Gyr, about 3.8 rotational periods after the encounter. }
\label{fig_xyz}
\end{figure}

\subsection{Simulations}

An extensive set of restricted n-body simulations guides our approximate evaluation of the linear gap rate, Equation~\ref{eq_Ratedef}, in a general context.   The background potential is an NFW density profile as fitted to the \citet{Aquarius} simulation of a Milky Way or M31 like galaxy. The stream is idealized as a circular ring at either 30 or 60 kpc of $10^4$ mass-less particles.  The ring of particles is initiated on tangential orbits at the local circular velocity, $v_c(r)=\sqrt{-\nabla \Phi \cdot r}$. That is, there are no random velocities, although the particles could be considered the guiding centers of a set of particles. Only one sub-halo is sent towards the ring in each simulation. An individual sub-halo is adequately modeled as a \citet{Hernquist} sphere with the mass-scale radius relation found in the Aquarius simulations, since for our purposes the exact details of the internal mass distribution are not important. The sub-halo motion is simply at a fixed velocity in a straight line, so it crosses the ring only once which avoids the complications of a sub-halo orbiting and having multiple ring crossings. A single sub-halo is started at a distance of the ring radius from the ring. The time step is 0.001 unit of $r/v_c(r)$ at 30 kpc, or about $1.4\times 10^5$ yr. The ring of particles responds to the sub-halo and background halo potential but has no self-gravity. A series of simulations is done varying the angle with respect to the tangent to the ring in the direction of motion over $\theta=$ [41, 60, 90, 104, 120, 139] degrees, and the angle with respect to the plane of the ring of $\phi=$ [0, 30, 60, 90, 270, 300, 330] degrees. Together these two distributions non-redundantly coarsely sample an isotropic velocity dispersion field. The encounter velocities are varied over  [0.3, 0.5, 0.8, 1.2, 1.6] times the circular velocity. For an isotropic velocity dispersion of sub-halos in our potential the typical ring encounter velocity is close to $v_c(r)$. The simulations are done for $10^6, 10^7, 10^8$ and $10^9\msun$ sub-halos. In total 9310 encounters are simulated for a ring.

The outcome of a typical set of  encounters after 6.92 Gyr is shown in Figure~\ref{fig_xyz}.  For purposes of easily visible illustration we chose a relatively massive sub-halo, $10^8 \msun$, which has a scale radius of 0.83 kpc. The sub-halo is moving at $1.2 v_c$.  The top two rows show impact parameters of 0.06 kpc, where the sub-halo significantly penetrates the ring. The bottom row shows an encounter with an impact parameter of 1.5 kpc, about twice the sub-halo scale radius. The outcome at late times in all cases is that the ring is primarily perturbed perpendicular to the plane with loops pulled out.  Note that  in Figure~\ref{fig_xyz} the vertical extent of the loops is stretched a factor of 220.  The ratio of the upward and downward loop heights depends on the angle $\phi$, as illustrated in the two upper rows.  The upward or downward extension also varies with time as the perturbed particles pass through the orbital plane. For the large $b$ relative to $R_s$ passage of  in the bottom row of Figure~\ref{fig_xyz}, the vertical deflection of the ring particles is somewhat larger than in the upper two rows, but there is essentially no visible gap in the ring density. The top two rows suggest that the density gap depth and length is insensitive to the angle $\phi$, the angle relative to the orbital plane. 

The gaps in these completely cold rings have a very simple morphology, whereas the gaps created in more realistic stellar streams with internal velocity dispersion and a finite width are much more complex \citep{M31_NW:11,YJH:11}. In particular real gaps are often slanted to the centerline. Old streams can be subject to so many encounters that the remaining segments of the stream are relatively dense clumps smaller than the gaps between. If the progenitor is on a fairly epicyclic orbit the tidal stripping episodes occur at perigalacticon and lead to additional structure in the stream. Therefore the simplified model presented here is intended to be a statistical guide to the analysis of streams, not a detailed description of any particular stream.

\begin{figure}
\begin{center}
\plotone{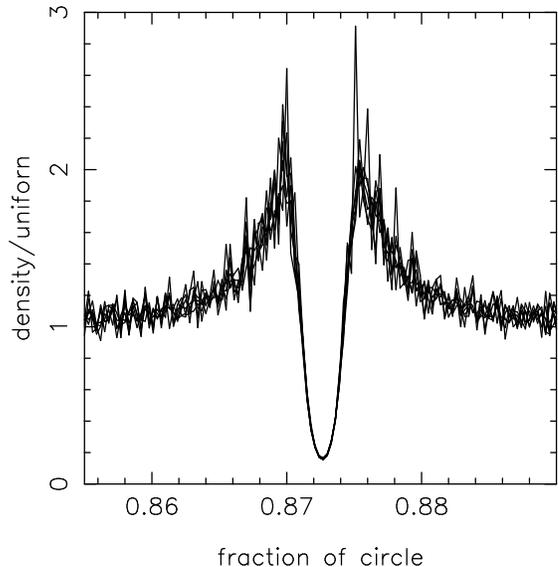}
\end{center}
\caption{The density distribution in the ring which a $10^7 \msun$ sub-halo passed at right angles in the horizontal plane at a velocity of $1.2v_c$ at $b=0.3$ kpc for the seven different angles, $\phi=$ [0\degr, 30\degr, 60\degr, 90\degr, 270\degr, 300\degr, 330\degr], with respect to the orbital plane.  }
\label{fig_denphi}
\end{figure}

\begin{figure}
\begin{center}
\plotone{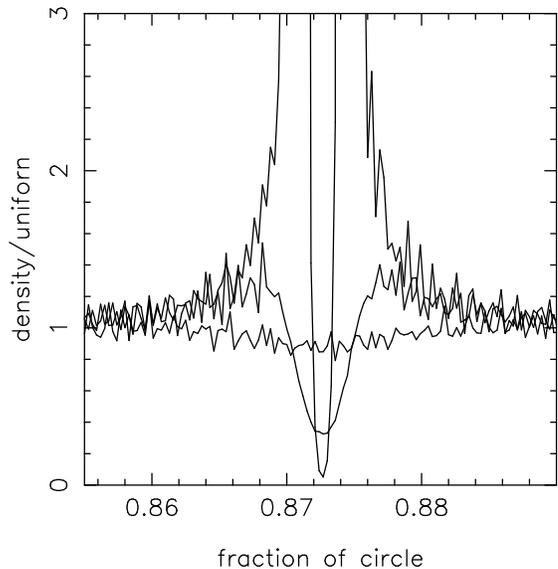}
\end{center}
\caption{The density distribution in the ring which a $10^7 \msun$ sub-halo passed at right angles in the horizontal plane at a velocity of 1.2 times the circular velocity with a closest separation distance of 0.06, 0.6 and 3 kpc, which lead to increasingly shallow density minima.}
\label{fig_den3}
\end{figure}

We measure the xy projected linear density relative to the unperturbed density using the distance, $\Delta d_{xy}$ between every third particle. The relative density is then $n\Delta d_{xy}/(6\pi r)$, where n is the number of particles in the ring and $r$ is its radius.   Figure~\ref{fig_denphi} shows that the density profile, of the gap is effectively independent of $\phi$, the encounter angle with respect to the plane. The particles that are stretched out in the loop pile up at the edges of the gap. The dependence of the depth of the density dip in the gap with $b$ is shown in Figure~\ref{fig_den3}, showing that although a loop is pulled out, there is little density dip for encounters significantly beyond the sub-halo scale radius. 

\subsubsection{Critical Impact Parameter}

\begin{figure}
\begin{center}
\plotone{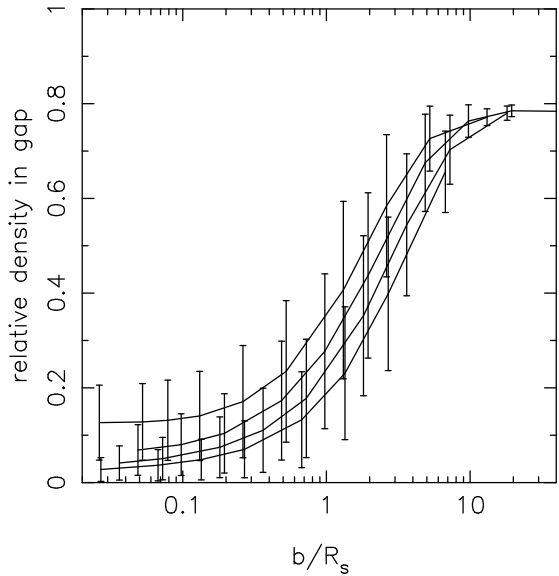}
\end{center}
\caption{The minimum density of particles in the plane of the ring as a function of the distance of closest approach normalized to the scale radius of the sub-halo. The lines are for sub-halo masses of $10^6, 10^7, 10^8$ and $10^9 \msun$ from top to bottom, respectively. The measurements are made at 6.6 Gyr after the encounters for $v=1.2v_c$. These measurements are made for 30 kpc radius rings.}
\label{fig_gapdensity}
\end{figure}

The relative density at the bottom of the dip is displayed as a function of the ratio of the impact parameter to the scale radius, $b/R_s(M)$, in Figure~\ref{fig_gapdensity}. The density measurements are averaged over all angles, $\theta$ and $\phi$.  The results are displayed for velocities of 0.8, 1.2 and 1.6 $v_c$. Since the ring particles are moving at the circular velocity there is a range of encounter velocities with the particles in the ring.  If the sub-halo velocity is significantly below the circular velocity there will be no very low velocity encounters in the frame of an individual star and the sub-halo, so on the average the gaps will be less deep. However since encounters below $0.8v_c$  comprise only about 10\% of the encounters we leave them out of the average. 

\begin{figure}
\begin{center}
\plotone{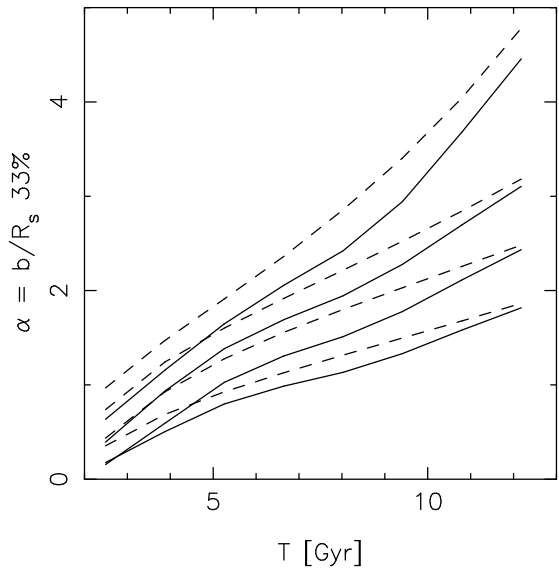}
\end{center}
\caption{The values  of $\alpha(M,r,t) \equiv b/R_s$, at $1/3$ of the unperturbed density, at 30 and 60 kpc, solid and dotted lines respectively, for  sub-halo masses of $10^6, 10^7, 10^8$ and $10^9 \msun$ from top to bottom, respectively.}
\label{fig_bRmt}
\end{figure}

We need to determine the maximum impact parameter that creates a visible gap.  We will model the behavior of the density dip as a function of $b/R_s(M)$ as a step function at a value $\alpha(M,r,t)$, with visible gaps created for $b\le \alpha R_s(M)$ and no gaps for larger $b$ values. We evaluate the value of $\alpha=b/R_s(M)$ at \onethird\ of the unperturbed density. The resulting critical values of $b/R_s(M)$ are function of both mass and time, as shown in Figure~\ref{fig_bRmt}.  Using a density dip of \onehalf\ yields larger values of $\alpha$, but its use turns out to give values within our overall modeling uncertainties.   The function $\alpha(M,r,t)$ has a significant time behavior. At early times,  roughly before 2 Gyr, the gaps are not well developed and we exclude them from the fit. Over a Hubble time the value of $\alpha$ nearly doubles. The time behavior is mainly relevant for dynamically young streams or the young part of streams that are still being created where we will expect the stream to be far less gapped than our average value.

 Fitting to the data of Figures \ref{fig_gapdensity} and \ref{fig_bRmt}, we find,
\begin{eqnarray}
\alpha_{33\%}(M,r,t) = ( 1.71 + [0.28 	+ 0.10 \log_{10}{M_8}] [t -7 ] )   \nonumber \\  
 \left( {r\over {30 {\rm kpc}}} \right)^{0.23}  M_8^{0.12},~~ 
\label{eq_b33}
\end{eqnarray}
where t is in Gyr and normally we will simply use the value at 7 Gyr as a reasonable average value for $\alpha(M,r)$.  Fits which exclude the mildly non-linear $10^9\msun$ sub-halos give nearly equal values.

\subsubsection{The Gap Creation Rate}

We now have the elements to evaluate the gap creation rate, Equation~\ref{eq_Ratedef}. The maximum impact parameter which creates a gap is $b_{max}= \alpha(M,r,t) R_s(M)$ where we will use the 7 Gyr value of $\alpha$, roughly the average over a Hubble time. The mean approach speed to the ring is the integral over  $v_\perp$ which is $\sqrt{2/\pi}\sigma\simeq 0.80\sigma$ for a Gaussian velocity distribution.  The rate integral then becomes an integral entirely in the mass distribution,
\begin{equation}
{\mathcal R}_\cup(\hat{M},r) = 0.80 \pi \sigma n(r) \int_{\hat{M}}^\infty N(M) \alpha_{33\%}(M,r) R_s(M) dM.
\end{equation}
The Aquarius simulations \citep{Aquarius} find that $R_s\simeq 0.83 M_8^{0.43}$ kpc and that the mass distribution function is $N(M) dM = 234 M_{8}^{-1.9} dM_{8}$. The mass function gives the total number of sub-halos contained within the volume, $V_{50}$, that has a mean interior density 50 times the critical density which has a radius 433 kpc \citep{Aquarius}. Putting these pieces together we find,
\begin{equation}
{\mathcal R}_\cup(\hat{M},r) = 490 \,\sigma V_{50}^{-1} {n(r)\over n_0} \int_{\hat{M}_8}^\infty {M_{8}}^{-1.9} \alpha_{33}(M_8){M_{8}}^{0.43} \, dM_{8},
\end{equation} 
where $\hat{M}_8\equiv\hat{M}/10^8 \msun$ is the smallest mass halo that creates a visible gap. The normalizing density in the $V_{50}$ of  the radial density variation, $n(r)$, of the sub-halo population volume is $n_0$.
The mass integral is dominated by the smaller masses so we have set the upper mass limit to infinity. Evaluating the integral, choosing a local $\sigma=120 \kms$ and converting to astronomical units, we find,
\begin{eqnarray}
{\mathcal R}_\cup(\hat{M},r) = 0.0066\, \left({r\over {100 {\rm kpc}}}\right)^{0.23}{n(r)/n_0\over {6}} {\sigma \over{120 \kms}} \nonumber \\ 
\hat{M}_{8}^{-0.35} \, {\rm kpc}^{-1} {\rm Gyr}^{-1}, ~~~
\label{eq_Rate}
\end{eqnarray}
where we have used the \citet{Aquarius} measurement for $n(r)$ (their Fig. 11) which finds that the sub-halo density at 100 kpc is 6 times the mean within the $V_{50}$ volume. The rate at which gaps are created increases as the minimum mass sub-halo that can create a noticeable gap goes down, i.e. $\mathcal{R}_\cup \propto {\hat{M}}^{-0.35}$. Halo-to-halo sub-halo number differences \citep{Chen:11} will lead to $\pm$50\% or so variation in the rate in a specific halo.

\subsection{Gap Loop Heights}

An important observable quantity is the length of the gaps.  We measure the length as the distance in the orbital plane between the two outside points at \onethird\ of the vertical maximum of a loop. The length of a gap should be related to the distance that the loop of ring particles is pulled out of the orbital plane, that is, as the vertical extension of the ring increases, so does the length of the visible gap. This means we first need to understand the dynamics of the vertical perturbations. The impact approximation and the tidal approximation are the two general approaches used to describe the interaction of a satellite, such as a sub-halo, with a set of particles, the stars in the stream. The impact approximation is best suited to calculating deflections of individual stars from the rapid passage of a satellite and the tidal approximation is best suited to problems involving the differential gravitational field across a finite sized object. Both approaches are useful for calculating the gravitational heating of the perturbed objects and at their root both have the same underlying gravitational dynamics. 

The impact approximation assumes that both a sub-halo and a star stream orbit can be approximated as straight lines and sets aside the possibility that orbital resonances play a significant role. A sub-halo of mass $M$ moving past a star at a relative velocity $v$ with a distance of closest approach $b$ will induce a velocity perpendicular to its direction of motion in the direction of closest approach of $\delta v_\perp = 2GM/bv$. Since many of the encounters partially intersect the stream the relevant mass will be $M(b)$, the mass enclosed within the impact parameter, not the total mass of a sub-halo.  For an orbit that is approximately circular at radius $r$ and taking $\delta v_\perp$ to be in the vertical direction, then conservation of angular momentum requires that the maximum orbital deviation out of the plane is $z_m =r \delta v_\perp/v_c$.

\begin{figure}
\begin{center}
\plotone{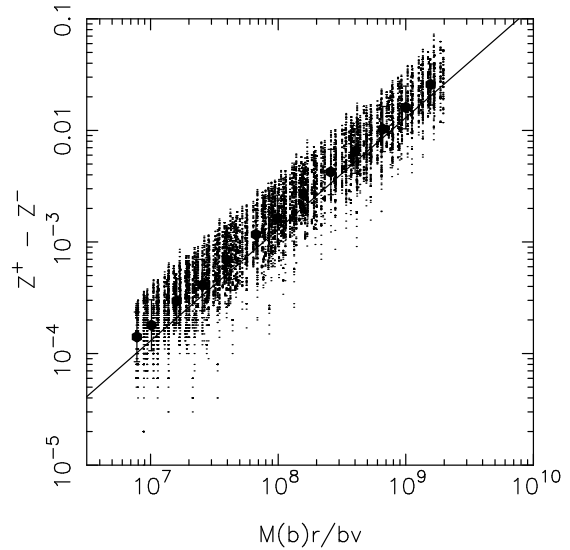}
\end{center}
\caption{The vertical extent at 33\% of maximum height of the sub-halo induced ``loop" to the impact parameter prediction for a ring at 60 kpc. The vertical extent and radial co-ordinate r are measured in scaled units of 30 kpc and the velocity in units of $v_c$. A ring at 30 kpc has half the vertical extent, as expected. }
\label{fig_binzmax}
\end{figure}

We need to average over sub-halos crossing  at all  angles to the ring tangent and to the orbital plane. To a first approximation forces in the horizontal plane do not add to the angular momentum around the normal to the plane of the orbit, but do act to coherently increase the epicyclic motion around the guiding center, but that makes little difference to the stream density. A sub-halo moving  at an angle to the orbital plane initially pulls the stream down and then up, which evolves into a paired set of peaks perpendicular to the orbital plane, as shown in Figure~\ref{fig_xyz}. The vertical extent of the perturbation is measured as the vertical distance from the positive peak to the negative peak, rather than the measurement of the absolute value of the offset from the initial orbital plane for a horizontal sub-halo passage. The resulting $Z^+-Z^-$ distribution is shown as a binned average in Figure~\ref{fig_binzmax}. The mean is about 30\% higher than the prediction, with about a factor of two scatter above and below the line. For our use the scale offset is not an issue.  

\begin{figure}
\begin{center}
\plotone{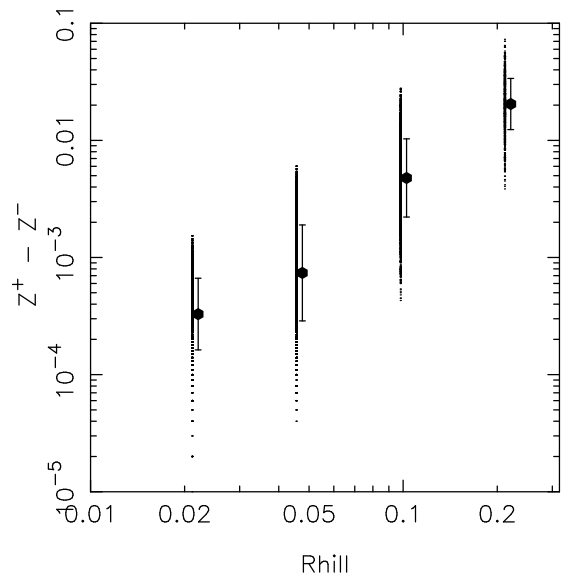}
\end{center}
\caption{The maximum vertical height of a perturbation as a function of the tidal Hill Radius. The averages are slightly offset for clarity.}
\label{fig_z_vs_Rhill}
\end{figure}

The tidal approximation is an alternative description of the encounter of a sub-halo with a stream.  The Hill radius, $R_{Hill}=r\sqrt[3]{M/(3M_t)}$, inside of which the gravitational field of the sub-halo, mass M, begins to dominate over the gravitational field of the total mass interior to the orbit of the stream, $M_t$. The plot of the vertical deflections against Hill radius is shown in Figure~\ref{fig_z_vs_Rhill} using the same data and same vertical scale as Figure~\ref{fig_binzmax}. Although the vertical perturbation and the Hill radius are strongly correlated there remains a larger scatter than with the impact approximation. Including a power of $b$ finds that $R_{Hill} b^{-1/3}$ removes some of the scatter. This is hardly surprising since with the addition of $v$ this becomes the impact approximation formula, multiplied by a constant. An alternative tidal quantity is the differential tidal field, which is proportional to $M/b^3$.  A plot of the vertical deflections against $M/b^3$ is mainly scatter. We conclude that the impact approximation is better at predicting the vertical deflections than the tidal approximation.

\subsubsection{The Characteristic Velocities}

Star streams have a mean particle motion close to the circular velocity. Therefore the encounter speed between a sub-halo and a star in the stream, which is required for the impact approximation, will on the average be higher than the mean circular velocity. A consequence is that low velocity encounters of sub-halos with the stream are extremely improbable. Taking the mean encounter along the stream to be $v_c$ and the velocity dispersion in the two axes perpendicular to the stream as $\sigma$ the typical relative velocity is $\sqrt{v_c^2+ 2\sigma^2}$. Using $\sigma\simeq v_c/1.8$ in the outer part of a galactic halo then the typical sub-halo star relative encounter velocity is $1.27v_c$. The full triple velocity integral has the complication that there is a singularity at relative velocity of zero. Physically this would correspond to the velocity of the stream's progenitor sub-halo. Inserting a very small ``core" velocity of about one meter per second, we find that the mean encounter velocity is $1.077v_c$. Combining these factors we find that on the average, 
\begin{equation}
z_m = {2GM r \over {1.08 v_c ^2b }}. 
\label{eq_impact}
\end{equation}

\subsubsection{Gap Lengths}

\begin{figure}
\begin{center}
\plotone{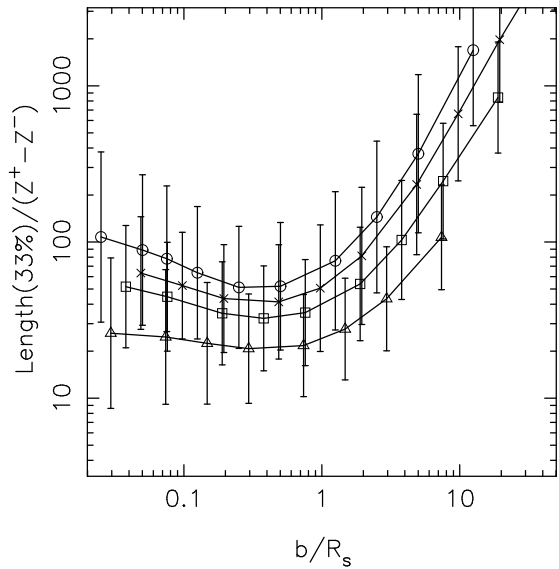}
\end{center}
\caption{The ratio of the linear size at 33\% of maximum height of the sub-halo induced ``loop" to the vertical height as a function of the impact parameter in ratio to the scale radius of the sub-halo. The lines show averages at fixed mass, starting with $10^6\msun$ at the top increasing to $10^9 \msun$ at the bottom. These values are for a ring at 60 kpc.}
\label{fig_binwidth}
\end{figure}

The length of a sub-halo induced gap as a function of the sub-halo parameters is needed to determine what gaps are large enough to be visible. We measure the gap length at ${1\over 3}$ of the maximum vertical height as a conservative approach. The basis of a relation is visible in Figure~\ref{fig_binwidth} in which gap lengths relative to $z_m$ are plotted for a ring at 60 kpc. We see that for substantially penetrating encounters, $b/R_s \le 2$, that the ratio of the gap length to the vertical extent, $z_m=Z^+-Z^-$,  is roughly a constant at a given mass.  Rings at smaller radii have relatively larger values of the length to height ratios, partly because $z_m$ includes an $r$ factor the ratio is dependent on the radius. 
Defining $\beta(M)$ as the length of the gap relative to the value of $z_m$ we can find an approximate fit for the mass and radial dependence over the range of 30 to 100 kpc to give, 
\begin{equation}
\beta_{33\%}(M,r) = 60 \left( {r\over{{\rm 30 kpc}}} \right)^{-0.63} M_8^{-0.11}.
\label{eq_w33}
\end{equation}
The function $\beta_{33\%}(M,r)$ appears to have no significant time dependence, once the gaps develop. At radii below 30 kpc the $\beta$ value becomes somewhat larger than Equation~\ref{eq_w33} gives.

The predicted length of a gap is the product of the average $\beta$ and the impact  Equation~\ref{eq_impact} for the vertical height of the loop in the star stream,
\begin{equation}
l(M) = \beta_{33\%}(M,r) {2 G M(R_s)r\over{ 1.08 R_s(M) v_c^2}}.
\end{equation}
Approximating a sub-halo density profile as a Hernquist profile we find that the mass inside the scale radius is $M(R_s)={1/4}M $ \citep{Hernquist}. Evaluating the numerical factors we find,
\begin{equation}
l(M,r) = {8.3 }\left( {r\over{{\rm 30 kpc}}}\right)^{0.37}M_8^{0.41}\,\kpc.
\end{equation}
We now have an equation that gives the typical gap length  as a function of sub-halo mass averaged over the velocity distribution. The gap length-mass relation can either be used as a consistency test, or, taken a step further and be used to eliminate the unobservable minimum mass parameter from the rate equation in favor of a minimum gap length.  

\subsection{The Gap Rate Stream Width Relation}

The width of the stream and the minimum visible gap size are linked through orbital dynamics. The stream width is largely the result of stars emerging through the tidal Lagrangian points with low initial velocity dispersion. The epicyclic approximation describes the subsequent motion of the stars with small velocity dispersions around a guiding center. An epicycle has comparable amplitudes in the direction of the motion and perpendicular to the motion, hence the stream width and the random motion along the stream are of nearly equal size. Provided that the phase angles of the motion are not significantly correlated, any induced gap narrower than the stream width will be blurred out. 
Therefore we make the assumption that the width of the stream is equal the minimum gap size and use it to estimate the minimum mass sub-halo that creates a gap,
\begin{equation}
w=l(\hat{M}) .
\label{eq_wdef}
\end{equation}
Solving Equation~\ref{eq_wdef} for $\hat{M}$ we substitute it into Equation~\ref{eq_Rate} to give the gap creation rate as a function of the width of stream, 
\begin{equation}
{\mathcal R}_\cup(w,r) = 0.059 \left( {n(r)/n_0\over{6}}\right) \left( {r\over{\rm 100\, kpc}}\right)^{0.55}
w^{-0.85} {\rm kpc}^{-1} {\rm Gyr}^{-1}, 
\label{eq_w_rate}
\end{equation}
for $w$ in kpc, where we have used Equations~\ref{eq_b33} and \ref{eq_w33}. 
Equation~\ref{eq_w_rate} recasts the gap rate in terms of an observable, assuming that the CDM halo substructure model gives the correct density of sub-halos. Over the range of 20-100 kpc  where streams are generally found the factors of $n(r)$ and $r$ partially counteract each other to reduce the radial variation.  Equation~\ref{eq_w_rate} is an observational test of the CDM sub-halo prediction. 

\section{Gaps in Observed Streams}

Counting gaps in a star stream requires considerable statistical care because the available observational data has streams embedded in a background which usually overwhelms the stream and is generally variable along the stream's length. A reliable measurement depends on having a good estimate of the density errors along the stream. The pseudo-images of stream star densities on the sky after filtering a star catalog with a color-magnitude relation 
can be usefully indicative but gaps should only be counted on extracted stream densities with local error estimates. We concentrate on four streams, the NW stream of M31, and the Pal~5, Eastern Banded Structure and Orphan streams of the Milky Way. All of these have published estimates of the linear density of stars along the stream. 
 
\subsection{The North-West Stream of M31}

\begin{figure}
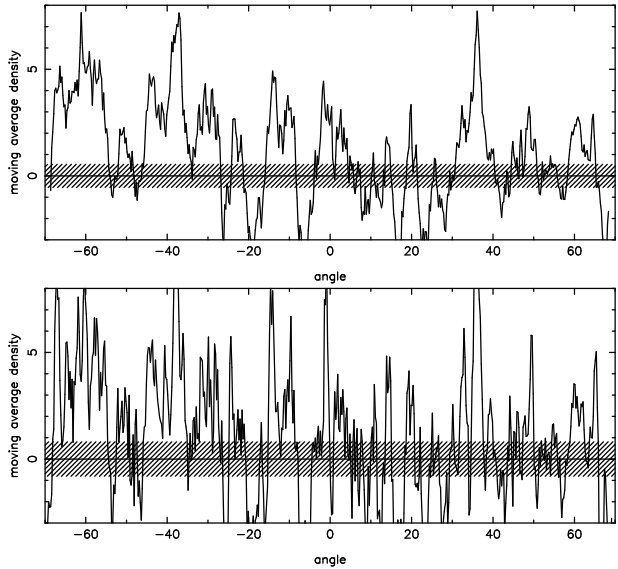

\begin{center}
\includegraphics{Fig9a.eps}
\includegraphics{Fig9b.eps}
\end{center}
\caption{The linear density of stars kpc$^{-2}$ in the M31 NW star stream with moving average of 5.5 kpc (top) and 2.5 kpc (bottom). The hatched areas are the $\pm 1\sigma$ average error of the mean. }
\label{fig_moving}
\end{figure}

\begin{figure} 
\begin{center} 
\plotone{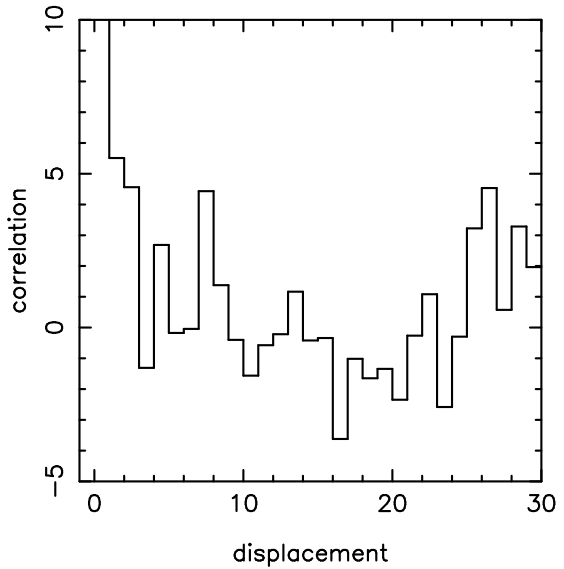}
\end{center}
\caption{The cross-correlation of the density field measured in 0.25\degr\ bins. The correlation is positive for the first 4 or 5 points, indicating a scale of about 1 degree. }
\label{fig_M31_dcor}
\end{figure}

The overall M31 PAndAS imaging project and the streams are described in \citet{Pandas,Richardson:11}  and the details of the NW stream in \citet{M31_NW:11}. Here we use the linear density profile to estimate the number of gaps. In Figure~\ref{fig_moving} we show the density field as a function of the angle around the star stream, with all quantities measured as reported in \citet{M31_NW:11}. The plots are made with a moving average over 5 and 11 bins of 0.25\degr, or roughly 2.5 and 5.5 kpc, respectively. To determine what scales in the stream appear to be physically correlated we plot the cross-correlation of the stream density minus the mean value with itself in Figure~\ref{fig_M31_dcor}. We see that the stream densities are positively correlated over the first three bins of 0.25\degr, after which the fluctuations are essentially random. This correlation length is comparable to the width of the stream and helps validate the assumption that leads to Equation~\ref{eq_wdef}. Therefore the smallest size scale we search for gaps in these data is about 1\degr, or, 2 kpc. 

To count the gaps we need to have a good measure of the level of statistical fluctuations. The cross-correlation of the data with itself, at shifts beyond 5 bins, where there is no intrinsic signal, provides an accurate internal estimate of the variance of the density field. The product of two normally distributed random variables is distributed like a modified Bessel function of the second kind and has a large kurtosis, 9, as compared to a normal distribution with a kurtosis of 3. However, the sum of many products (115 in our 0.25\degr\ sample) converges to a Gaussian distribution in accord with the central limit theorem. The kurtosis of $n$ sums of products is $3+6/n$ so declines fairly quickly to the  Gaussian value of 3 with increasing $n$. The variance around the mean density of the data within the stream region, prior to background subtraction, is 1.22 (stars/kpc)$^2$, and a comparable 1.33 in the background region. The background subtracted mean stream density has a variance of 6.16 (stars/kpc)$^2$. The F ratio of the measured variance over the randomly expected variance is 6.16/(1.22+1.33)=2.42, which has a random chance of occurrence of less than $2\times 10^{-5}$. Larger bins give even higher significance that the measured stream density has variations well in excess of what is expected from random fluctuations. The conclusion and the levels of significance are in accord with the results presented in \citet{M31_NW:11} but use an independent line of argument which does not rely on an explicit calculation of shot noise statistics.

An estimate of the number of gaps is simply to count the number of times the density crosses down and up through some density threshold, requiring that the gap be at least the size of the smooth length. For a 5 point moving average at [$-1\sigma,0,+1\sigma$] density levels we find [11,11,15] gaps of mean length [7.1,6.9,15.1] bins, or, [1.8, 1.7, 3.8] kpc. 
Taking the stream length to be 200 kpc and assuming an age of 10 Gyr we find from the $12\pm2$ gaps that the M31 NW stream has been subject to a mean gap creation rate of $0.006\pm50\%$ gaps kpc$^{-1}$ Gyr$^{-1}$. In principle the predicted rate also includes any sufficiently massive object, but the detectable mass is well above the range for globular clusters and we do not expect molecular clouds along the large radius orbit of the NW stream. 

The rate equation, Equation~\ref{eq_Rate}, applied to the M31 NW stream requires a minimum sub-halo mass of $\hat{M}=1.3\times 10^8\msun$ and implies a total of 205 sub-halos, which is a factor of 7.3 higher than 28 known dwarf galaxies of M31 \citep{Richardson:11} but includes objects out to 433 kpc, of which the only the inner 150 kpc radius volume has been carefully searched.  Therefore the M31 stream not demand a huge dark matter sub-halo population by itself, but this analysis does recover the well known discrepancy which begins to appear for even fairly massive sub-halos.

\subsection{The Pal~5 Stream}

A narrow stream is predicted to have more visible gaps relative to wider stream at larger radius. The globular cluster Pal~5 has both leading and trailing streams with the northern, trailing, arm having the most data in the Sloan survey. 
With data made available before DR1 \citet{Odenkirchen:03} found the stream to be about 6 degrees long and measured the width to be 0.11 kpc. With the better coverage of the SDSS DR4, \citet{GD:Pal5} found that the northern arm was about 18.5\degr\ and provided a surface density contour plot showing the considerable variation in the surface density of the stream. The density variations closest to the cluster are convincingly explained as epicyclic pileups of stars tidally stripped from the cluster \citet{Kupper:10,Kupper:11}, however beyond about 1kpc (2.5 degrees) the variation in phase angles smooths out this behavior. From the published Figure~3 of \citet{GD:Pal5} we count regions in the stream below the lowest contour to find  that there are about 5 gaps over the outer 16\degr\ of the northern tail of Pal~5. \citet{Odenkirchen:03} estimated that the 5-6\degr\ of stream took about 2 Gyr to fill out, which scaled to the 18.5\degr\ length gives an age of about 7 Gyr. With these values we estimate the linear gap creation rate in the northern stream of Pal~5 to be about 0.11 kpc$^{-1}$ Gyr$^{-1}$ to which we assign an error of 50\% based on uncertainties in the gap counts and stream age.

\subsection{The EBS Stream}

\citet{G11:EBS} finds that the Eastern Banded Structure is 0.17 kpc wide and located at a galactocentric distance of about 15 kpc. Figure~4 of his paper plots the density along the 18\degr\ of the stream, with an estimate error of 30\% in each bin of 0.5\degr. We estimate that there are between about 4 and 12 gaps along the stream. We will take the stream age to be approximately 7 Gyr to derive a gap creation rate of $0.24 $ gaps kpc$^{-1}$ Gyr$^{-1}$, with about a factor of two uncertainty in this value. 

The estimated rate of gap creation in the EBS stream with Equation~\ref{eq_Rate} implies a minimum halo mass of $\hat{M}\simeq 1\times 10^5 \msun$. The derived minimum mass implies a total of  $1\times 10^5$ sub-halos, although we note that both the mass and in turn the implied total number of sub-halos are sensitive to the gap rate estimate and the details of the model in Equation~\ref{eq_Rate}. Possible complications for the EBS stream is that  the sub-halos are subject to erosion by the disk and bulge \citep{DOnghia:10} and the stream may encounter smaller scale structures such as spiral arms and molecular clouds which could add to gaps in the stream.

\subsection{The Orphan Stream}

The Orphan Stream \citep{Grillmair:06,Belokurov:06,Belokurov:07} has received considerable attention,  culminating with a detailed orbital model \citep{Newberg:10}.  Increasing the mean distance from 20 to 30 kpc revises \citet{Grillmair:06}'s estimated width of 700 pc  to 1 kpc. Gaps are marginally present in Figure~5 of \citet{Newberg:10} at -16\degr\ and -31\degr, particularly when compared to the bottom panel of their Figure~17.  We estimate that there are 2 gaps present, although we will allow a 100\% error on this estimate.  \citet{Newberg:10} present a model that estimates the age of the stream to be about 3.9~Gyr.  If we had used the dynamical age of 3.9 Gyr rather than a uniform 7 Gyr the correction would raise the Orphan data into better agreement with the prediction.

\begin{deluxetable}{rr rrr rr}
\tablecolumns{7}
\tablewidth{0pc}
\tablecaption{Observed Stream Gap Statistics}
\tablehead{
\colhead{Stream} & \colhead{Gaps}   & \colhead{Length} & \colhead{Width}   & \colhead{Age} & \colhead{$R_{GC}$}& \colhead{$n/n_0$}\\
 & \# & [kpc] & [kpc] & [Gyr]  &[kpc] &  \\
}
 \startdata 
M31-NW 	& 12 	& 200 	& 5 		& 10	& 100	&6	\\
Pal~5		& 5		& 6.5	& 0.11	& 7		& 19	&24	\\
EBS			& 8		& 4.7	& 0.17	& 7		& 15 	&30	\\
Orphan 		& 2 		& 30 	&1.0 	& 3.9	& 30	&20 \\
\enddata
\end{deluxetable}

\begin{figure} 
\begin{center} 
\plotone{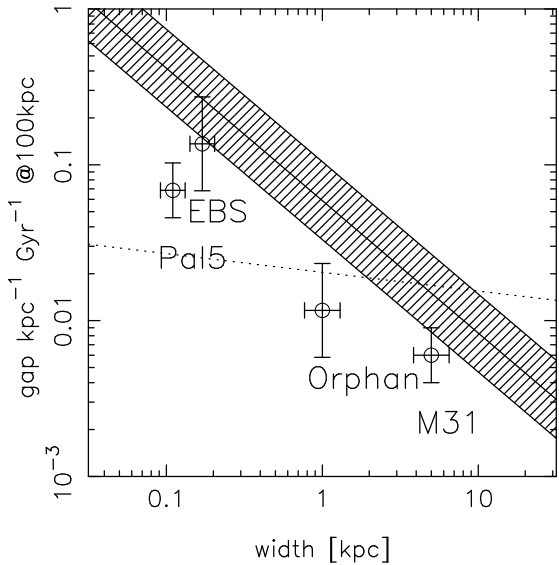}
\end{center}
\caption{The estimated gap rate vs stream width relation for M31 NW, Pal~5, the EBS and the CDM halo prediction.  All data have been normalized to 100 kpc. The width of the theoretical relation is evaluated from the dispersion in the length-height relation of Fig.~\ref{fig_binwidth}. Predictions for an arbitrary alternative mass functions, $N(M)\propto M^{-1.6}$ normalized to have 33 halos above $10^9\msun$ is shown with a dotted line.}
\label{fig_gaprate}
\end{figure}

\section{Comparison of Predictions and Observations}

The observed stream gap rate and width estimates are summarized in Table~1. The results of this paper are largely encapsulated in Figure~\ref{fig_gaprate}, which plots the measured and predicted gap rates against the widths of the streams. The data have all been rescaled to a galactocentric distance of 100 kpc using $n(r)/n_0$ and $r$ in Equation~\ref{eq_w_rate}. The agreement of the data and the CDM sub-halo prediction is remarkable. We show the relation for an arbitrary shallower $N(M)\propto M^{-1.6}$  normalized to have the same number, 33, sub-halos above $10^9\msun$ present in the LCDM prediction. If one used the visible dwarf galaxies to predict the number of sub-halos the predicted density of gaps in these streams would be roughly a factor of 100 smaller, which would be in wild disagreement with the observation of virtually any gaps at all. The numerical consistency that we find is both an indirect and statistical test for the existence of the large predicted population of dark matter sub-halos.

The Orphan stream merits further observational attention since the available measurements are potentially consistent with it being a stream with no gaps. There is room for variation in these statistical estimates, but when the other streams work fairly well with the adopted CDM spectrum the Orphan stream is expected to fall into line and could potentially be a case against a rich sub-halo population in our galaxy. 

\section{Discussion and Conclusions}

This paper lays a basis for understanding the rate at which dark matter sub-halos create visible gaps in stellar streams. Restricted n-body experiments guide a calculation based on the impact approximation. The outcome is a prediction of the local rate at which gaps are induced in a stellar stream which effectively counts the number of sub-halos massive enough to produce visible gaps. The rate of gap creation is cast as a relationship with the stream width, which gives a predicted relationship between observables that tests the CDM sub-halo prediction. 

Comparison of the CDM based prediction of the gap rate-width relation with published data for four streams shows generally  good agreement within the fairly large measurement errors. The result is a statistical argument that the vast predicted population of sub-halos is indeed present in the halos of galaxies like M31 and the Milky Way. The data do tend to be somewhat below the prediction at most points. This could be the result of many factors, such as the total population of sub-halos is expected to vary significantly from galaxy to galaxy, allowing for the stream age would lower the predicted number of gaps for the Orphan stream and possibly others as well, and most importantly these are idealized stream models.

There are many improvements possible to strengthen, or, destroy this result.  Each observed stream could be realized through an n-body model and inserted into an appropriate model of a sub-halo rich galactic halo. The result would still be statistical but the various time dependent and mass dependencies would be fully taken into account. On the observational side the presence or absence of a particular gap in a stream are somewhat uncertain due to the relatively low signal to noise of streams at the present time. A statistically robust measure of the number of gaps needs to be developed. However, the observations are improving quickly and much more secure gap descriptions will soon be available.  Overall, within our simplified general modeling of sub-halos crossing streams to create gaps, we finding a perhaps surprisingly good agreement between data acquired for other purposes and the CDM prediction of a large population of sub-halos.

\acknowledgements
This research is supported by NSERC and CIfAR.  Rosie Wyse, Carl Grillmair and an anonymous referee made helpful suggestions.

\end{document}